\documentclass[conference,letterpaper,10pt]{IEEEtran}

\usepackage{multirow}
\usepackage{graphicx,comment} 
\usepackage{tabularx}
\usepackage{amsmath,cases}
\usepackage{url}
\usepackage{color}
\usepackage{epstopdf}
\usepackage{enumitem}
\usepackage{dsfont,amssymb}
\usepackage[algoruled,medskip,dontprintsemicolon,linesnumbered,Algorithm]{algorithm}
\usepackage[noend]{algpseudocode}

\newcommand{\Fig}[1]{Fig.~\ref{fig:#1}}
\newcommand{\Sec}[1]{Sec.~\ref{sec:#1}}

\newcommand{\Eq}[1]{(\ref{eq:#1})}
\newcommand{\Alg}[1]{Alg.~\ref{alg:#1}}
\newcommand{\Line}[1]{Line~\ref{line:#1}}

\newtheorem{example}{Example}

\newcommand{\Qc}{\mathcal{Q}}
\newcommand{\Kc}{\mathcal{K}}

\newcommand{\Wc}{\mathcal{W}}

\newcommand{\PP}{\mathds{P}} 

\def\BibTeX{{\rm B\kern-.05em{\sc i\kern-.025em b}\kern-.08em
    T\kern-.1667em\lower.7ex\hbox{E}\kern-.125emX}}
\begin{document}
\setlist{nolistsep,noitemsep,leftmargin=*}

\title{An Edge-powered Approach to Assisted Driving}

\author{Francesco Malandrino\textsuperscript{a, c}, Carla Fabiana Chiasserini\textsuperscript{b, a, c}, Gian Michele Dell'Aera\textsuperscript{d}\\
a: CNR-IEIIT, Italy -- b: Politecnico di Torino, Italy -- c: CNIT, Italy -- d: TIM, Italy
}

\maketitle

\begin{abstract}
Automotive services for connected vehicles are one of the main
fields of application for  new-generation  mobile networks as
well as for the edge computing paradigm. In this paper, we investigate a system
architecture that integrates the distributed vehicular network with
the network edge, with the aim to 
optimize the vehicle travel times. We then  present a queue-based system model
that permits the optimization of the vehicle flows, and we show its
applicability to two relevant services, namely, 
lane change/merge (representative of cooperative assisted driving) and navigation.  
Furthermore, we  introduce an efficient algorithm called Bottleneck
Hunting (BH), able to formulate high-quality flow policies in linear time.
We assess the performance of the proposed system architecture and of
BH through a comprehensive and realistic simulation framework,
combining ns-3 and SUMO. 
The results, derived under real-world scenarios, show that our solution provides much shorter travel times than
when decisions are made by individual vehicles.  
\end{abstract}

\section{Introduction}
\label{sec:intro}

Vehicle traffic congestion has been identified as a prime cause of road accidents, with their significant cost in term of time and lives~\cite{who-roads}. Among the strategies to alleviate congestion, {\em coordination} between vehicles has been studied in several contexts, including lane maneuvers~\cite{zhou2016impact}, intersection management~\cite{lehoczky1972traffic}, and traffic light optimization~\cite{dunne1967traffic}. The emergence of connected vehicles, with their assisted- and self-driving capabilities, has fostered further interest in this topic, as witnessed by European projects~\cite{autonet30-commag,5gcar,5gt} and international standards~\cite{etsi-2019}.

\begin{figure}[t]
\centering
\includegraphics[width=1\columnwidth]{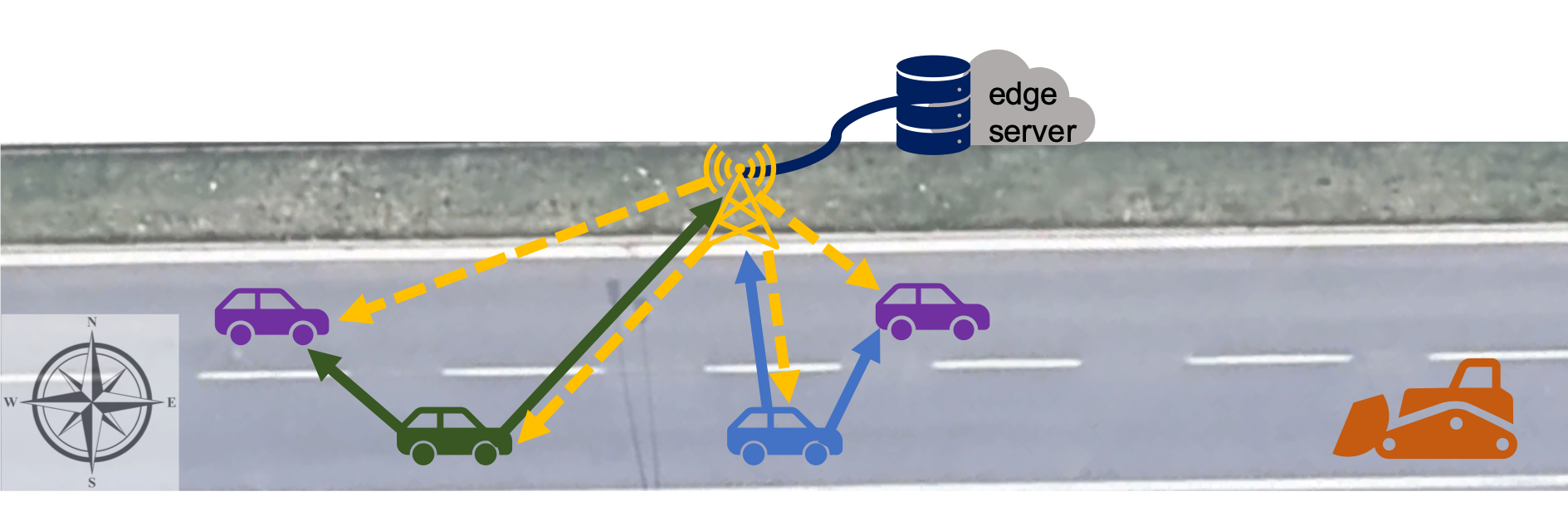}
\caption{Lane change:  the south lane is blocked  and
the two vehicles using it must move to the north lane. Leveraging the
messages sent by 
vehicles (solid arrows), an edge server defines the optimal
policy  and sends it to a road-side actualizer. 
The latter translates it into individual instructions, transmitted to
each vehicle  (dotted arrows). 
\label{fig:archi}
\vspace{-5mm}
} 
\end{figure}

However, present-day research efforts envision that vehicles make {\em local} decisions, and aim at providing such vehicles with as much information as possible. In view of the fact that local decisions have far-reaching consequences on the global traffic flow, we envision an alternative system architecture, based on the edge computing paradigm and summarized in \Fig{archi}, where:
\begin{itemize}
    \item an edge server formulates {\em policies}, concerning all traffic flows;
    \item actualizers located at the infrastructure, e.g., cellular eNBs or road-side units (RSUs) translate such policies into local, per-vehicle {\em decisions}.
\end{itemize}
With reference to \Fig{archi}, a policy may dictate that 60\% of vehicle traveling on the south lane should turn left early (where the green vehicle is), and the rest later (where the blue vehicle is). Given this policy, actualizers can give advice to individual vehicles, chosen randomly according to a 60\%-40\% proportion, on when to change lanes.

Formulating the policies at the edge server is not an easy task, for
several reasons. First, policies need to account for how vehicles with
different destinations sharing the same stretch of road influence each
other's travel times. Second, in order to ensure fairness, it is not
enough to optimize the {\em average} travel time of vehicles with a
given source and destination, rather the travel time distribution over
all vehicles needs to be considered. Third, all such decisions must be
made effectively 
-- optimally, if possible -- and efficiently, in order to quickly
react to changes in the vehicle traffic.

In this paper, we address the task of formulating policies at the edge server, by making the following  main contributions:
\begin{itemize}
    \item introducing a system architecture, detailed in \Sec{archi},
      combining the distributed vehicular network with an 
      edge-based decision-maker, whose task is to formulate traffic
      flow-level policies;
    \item proposing, in \Sec{model}, a queue model 
      describing arbitrary road topologies, the paths taken by
      vehicles, and the distribution of their travel times, in a
      tractable manner; 
    \item through the above model, deriving the distribution of the vehicles' travel times, in \Sec{times}, and formulating the problem of optimizing it, in \Sec{problem};
    \item presenting, in \Sec{algo}, an optimal algorithm with linear
      complexity, called
      {\em Bottleneck Hunting} (BH), along with a discussion on its novelty with respect
      to traditional multi-path packet routing.
\end{itemize}
The performance evaluation results, obtained through the methodology outlined in \Sec{validation} and summarized in
in \Sec{results}, confirm that the  policies formulated by BH result
in much shorter travel times than the decisions made by individual
vehicles. 
Finally, after reviewing related work in \Sec{relwork}, we conclude the paper in \Sec{conclusion}.

\section{System architecture}
\label{sec:archi}

We now introduce our new architecture, highlighting how it is
consistent with existing standards on
connected vehicle communications.  
According to the ETSI 102.941 (2019) standard, connected vehicles
periodically (e.g., every 100 ms) broadcast cooperative awareness messages (CAMs),
including their location, speed, and heading. Such information is
then used by other vehicles and/or the network infrastructure for
several safety and convenience 
services, e.g., \cite{noi-vtm}. Upon detecting a
situation warranting action, decentralized environmental
  notification messages (DENMs)  are sent to the affected
vehicles, so that their actuator is triggered and/or their driver
is warned. Both CAMs and DENMs can carry additional information~\cite{etsi-2019} for the support of  assisted driving services such
as lane change/merge \cite{etsi-2019} and navigation services (see ETSI  102.638). 

None of the above is envisioned to change under our proposed
architecture; in particular, all services still leverage the transmission of 
CAMs and DENMs.  For concreteness, we focus on  the lane change assistance
and navigation services.  
In the first case, the edge server exploits
the information in the CAMs received by the radio access nodes (and
possibly notifications of lane blocks carried by DENMs) to formulate  an
optimal policy. Such a policy is then sent to the actualizers, which,
being closer to the mobile users, can account for the most recent CAMs
and exploit them to  translate the policy  into individual
instructions for the single vehicles (see \Fig{archi}). These instructions are notified
to vehicles through DENMs sent by the radio access nodes. Upon
receiving a DENM, a vehicle starts interacting via V2V communications with its neighbors
following, e.g., the protocol defined in \cite{autonet30-commag} and
using CAMs and DENMs as foreseen by \cite{etsi-2019}.  

The second service (navigation) operates on a longer time and geographical  scale: the edge server exploits the CAMs
and, in particular, the route destination field therein, and computes
the optimal vehicles' route. This is then notified by the radio
access nodes using DENMs.

\section{Model design}
\label{sec:model}

In this section, we describe how our system model represents the road
layout, the vehicle flows (\Sec{sub-queues}), the routes taken by
vehicles in the same flow, and their travel times
(\Sec{sub-paths}). 
Importantly, 
\begin{itemize}
    \item based on the existing works and validation studies (see \Sec{relwork}), we consider Markovian arrivals, of either individual vehicles or  batches thereof;
    \item we do not restrict ourselves to a specific service time distribution, i.e., consider generic M$^X$/G/1 queues;
    \item given the scope of our work, we focus on  uncongested
      scenarios where the number of vehicles traveling on a road stretch
      is lower than its capacity, hence road stretches can be modeled with queues with
      infinite length. 
\end{itemize}

The decisions to be made correspond, intuitively, to the policies
formulated 
by the edge server in  \Fig{archi}. Specifically, they
consist of  the suggested travel speed  and the probabilities of
taking a given lane/stretch of road. In the following,
we refer to a single lane on a stretch of road as {\em road segment}.

\begin{figure}
\centering
\includegraphics[width=0.8\columnwidth]{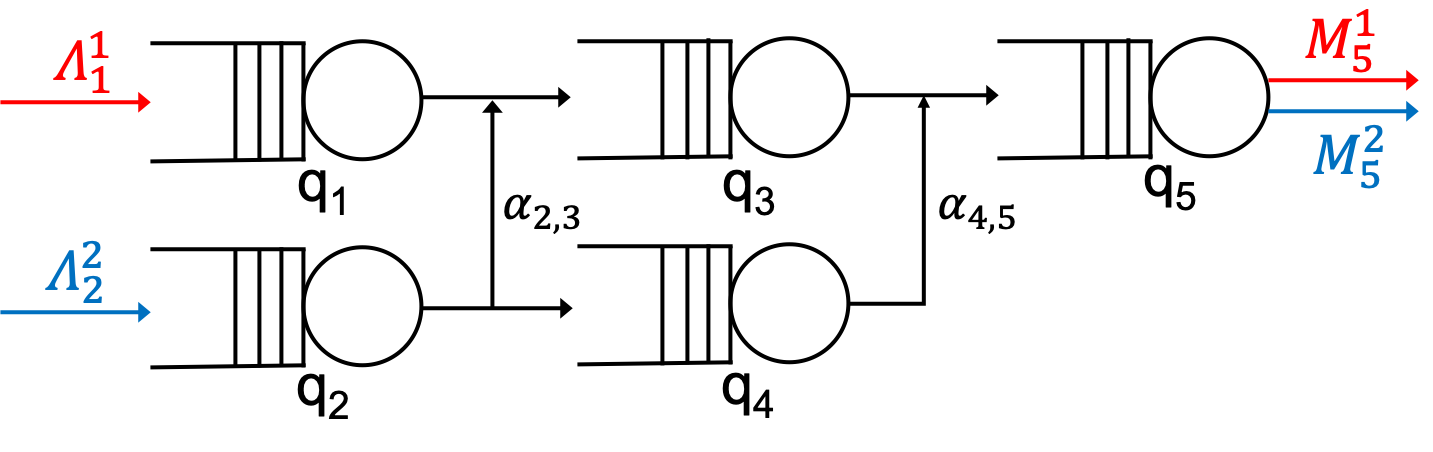}
\caption{Toy scenario: flows of vehicles go through a set of interconnected queues.\label{fig:scenario}
\vspace{-5mm}
} 
\end{figure}

\subsection{Road topologies as queue networks}
\label{sec:sub-queues}

We model the road topology under the control of the edge server as a set of interconnected 
  queues~$q_i\in\Qc$, as exemplified in \Fig{scenario}.  
Similarly to existing works (including experimental validation, see \Sec{relwork}), 
each queue represents a road segment, characterized, e.g., by a certain speed
limit and capacity.

Service rates,  $\mu_i$, are bounded by a maximum
value~$\mu^{\max}_i$, determined by the length of the road
segment modelled by queue $q_i$, its speed limit, and the
inter-vehicle safety distance: 
\begin{equation}
\label{eq:mu-max}
\mu_i\leq\mu^{\max}_i\quad\forall q_i\in\Qc \,.
\end{equation}

Several traffic {\em flows}, $k\in\Kc$, may
travel across the road topology or part of it, each flow being defined as a set of vehicles having the same source and
destination within the road topology. 
Specifically, given a flow $k$ entering and leaving the road topology
at queue $q_i$ and $q_o$, respectively, the parameter~$\Lambda^k_i$
denotes the vehicles 
arrival rate, 
while parameter~$M^k_o$ denotes the rate at which they
exit the  topology. 
Notice that, for every flow~$k$, there will be
exactly one queue for which~$\Lambda^k_i>0$ and one for
which~$M^k_o>0$ (see \Fig{scenario}).
Vehicles move from a generic queue $i$ to a queue $j$ according to
probability~$\alpha_{ij}$. 

Given the above model, our decision variables are the service
rates~$\mu_i$ of queues $q_i\in\Qc$, and the transition
probabilities~$\alpha_{i,j}$ 
(although some $\alpha_{ij}$ can be fixed, e.g., in \Fig{scenario},
$\alpha_{4,5}=1$). 
The total incoming flow for queue~$q_i$ is denoted with~$\lambda_i$
and depends on $\Lambda_i^k$, $M_o^k$, $\alpha_{i,j}$, and the probability~$\pi^0_i$ that queue~$q_i$ is empty:
\begin{equation}
\label{eq:lambda-i-generic}
\lambda_i= \sum_{k\in\Kc}\Lambda^k_i
+\sum_{h\in\Qc}\alpha_{h,i}\left(\mu_h(1-\pi^0_h)-\sum_{k\in\Kc}M^k_h\right).
\end{equation}
Eq.\,\Eq{lambda-i-generic} can be read as follows: the flow incoming
into queue~$q_i$ is equal to the flow of vehicles that begin their
journey therein, plus a fraction~$\alpha_{h,i}$ of the vehicles that
exit other queues~$q_h$, but do not leave yet the road topology.

\subsection{A path-based view} 
\label{sec:sub-paths}

Vehicles of the same flow, i.e., having the same source and the same
destination within the road topology, may nonetheless have different
trajectories, hence, traverse different sequences of queues. 
We define such sequences of queues as {\em paths}~$w\in\Wc$.
Each path  $w$  is an array including the queues traversed
by vehicles taking it; given a queue~$q_i\in\Qc$, we  
write~$q_i\in w$ if  path~$w$ includes~$q_i$. 
The probability that vehicles of flow~$k$ take
path~$w$ is denoted by $p_w^k$; each path~$w$ is used by one flow~$k$ only, indicated
as~$\kappa(w)\in\Kc$. 
Introducing paths
allows for a flexible relationship between flows and
queues, thus enhancing the realism of our model, while keeping its
computational complexity low.

Importantly, given  the edge-based policy, 
vehicles are assigned a path (according to the $p_w^k$~probabilities)
at the beginning of their travel across the considered road topology, 
before they enter the first queue. Thus, given the path it takes, the
queues traversed by a vehicle are unequivocally determined.

The probabilities~$p_w^k$ that vehicles of flow~$k$ take path~$w$
can be expressed as a function of the probabilities~$\alpha_{i,j}$ as follows:
\begin{equation}
\label{eq:pwk}
p_w^k=\begin{cases}
\prod_{n=2}^{|w|}\alpha_{w[n-1],w[n]} & \mbox{ if } \kappa(w)=k\\
0 & \text{otherwise,}
\end{cases}
\end{equation}
where~$w[n]\in\Qc$ is the $n$-th queue included in path~$w$.
Clearly, the $p_w^k$~probabilities have to sum up to one:
\begin{equation}
\label{eq:pwk-sum-one}
\sum_{w\in\Wc}p_w^k=1,\quad\forall k\in\Kc.
\end{equation}

We  stress that, while each path is unequivocally
associated with one flow, the opposite does not hold. 
With reference to \Fig{scenario}, flow~$1$ is
associated with one path~$[q_1,q_2,q_5]$ only, while 
flow~$2$ can use path~$[q_2,q_3,q_5]$
or~$[q_2,q_4,q_5]$. 
As for paths and queues, there is a many-to-many relationship between
them. In \Fig{scenario}, 
queues~$q_1$ and~$q_4$ belong to one path each, queues~$q_2$ and~$q_3$
to two paths each, and queue~$q_5$ to all three paths.

When considering paths, the local arrival rate at each queue $q_i$ 
in \Eq{lambda-i-generic} depends on: 
the paths $q_i$ belongs to, the probability that flows take
those paths, and the arrival rate of those flows. Specifically,
\begin{equation}
\label{eq:lambda-p}
\lambda_i=\sum_{k\in\Kc}\Lambda^k\sum_{w\in\Wc\colon q_i\in w}p^k_w.
\end{equation}

\section{Path Travel Times} 
\label{sec:times}

In the following, we leverage the path-based view of the system to
characterize the travel time experienced by each flow~$k$. 
Let us first denote with $f_i(t)$ the probability density function (pdf) of the sojourn time at an
individual queue~$q_i\in\Qc$.  Such pdf is a function of $\lambda_i$ and
$\mu_i$, according to an expression that depends on the statistics of
the queue arrival process and the service time.
Then,  recalling that vehicles taking path~$w$ traverse all queues in $w$,  the travel time associated with path~$w$ is the sum of the sojourn times at all queues therein. The pdf of such a time is the $|w|$-way convolution of the individual pdfs associated with each queue, i.e.,
$f_w(t)=\textsf{Conv}_{q_i\in w}f_i(t)$.
By integrating $f_w(t)$, we compute the cumulative density functions
(CDFs) and take their 
Laplace transform, thus obtaining:
\begin{equation}
\label{eq:cdf-s-generic}
\mathbf{F}_w(s)=\frac{1}{s} \prod_{q_i\in w} \mathbf{f}_i(s),
\end{equation}
where $\mathbf{f}(s)=\mathcal{L}[f(t)](s)$ is the Laplace transform
of~$f(t)$. 

Anti-transforming, we can  obtain the CDF of the path-wise travel time:
$F_w(t)=\mathcal{L}^{-1}\left[\mathbf{F}_w(s)\right](t)$.
We recall that~$F_w(t)$ is a function of the control
variables~$\mu_i$ and~$p_w^k$ (i.e., $\alpha_{h,i}$), which appear in the pdf~$f_i(t)$. As
mentioned, the
actual form of~$f_w(t)$ and~$F_w(t)$, as well as of the above Laplace
transforms, depends on the queue arrival process and service time
distribution. In  many cases of
interest, $\mathbf{F}_w(s)$ can be expressed as a ratio
between polynomials, and anti-transformed into a summation of terms of
type~$At^n\mathrm{e}^{t\tau}$.

Next, we  move from paths to flows. To this end,
$F_w(t)$ can be exploited to write the
probability~$\delta_w(\hat{t})$ that the travel time of a vehicle
taking path~$w$ 
exceeds a value~$\hat{t}$:
\begin{equation}
\label{eq:prob-w}
\PP(\text{travel time on path $w$} > \hat{t})=1-F_w(t)|_{t=\hat{t}}
\triangleq \delta_w(\hat{t}).
\end{equation}
We now need a flow-wise equivalent of $\delta_w(\hat{t})$. This  
however cannot be computed using the CDF
of the per-flow travel time, i.e., similarly to  \Eq{prob-w}, because such a distribution, in
general, is {\em not} a linear combination of the path-wise CDFs.
 Instead, we proceed as follows. 

Let~$\omega^k$ be a flow-wise target travel time, 
e.g., the ratio of the distance between the flow source and
destination to the desired average speed between them. Given
the $\omega^k$~value, a good measure of the flow quality of
service is given by the probability~$\delta^k(\omega^k)$ that the
travel time of the  vehicles
of flow~$k$ exceeds~$\omega^k$. To express such
values, we 
can combine the $\delta_w$ path-wise probabilities in \Eq{prob-w} with
the values~$p_w^k$, expressing the probability that a vehicle of flow~$k$ takes path~$w$, and write:
\begin{equation}
\label{eq:prob-k}
\PP(\text{travel time of  flow $k$} \mathord{>}
  \omega^k)= \sum_{w\in\Wc}p^k_w\delta_w(\omega^k) 
 \triangleq  \delta^k(\omega^k) \,.
\end{equation}

\section{Problem formulation}
\label{sec:problem}

Our high-level goal is 
to keep the fraction of vehicles of each flow~$k$, whose travel time
exceeds the target~$\omega^k$, as low as possible. 
To ensure fairness among vehicles of different flows, we
formulate such a goal as the following min-max objective:
\begin{equation}
\label{eq:obj-generic}
\min_{p^w_k,\mu_i}\max_{k\in\Kc}\sum_{w\in\Wc}p^k_w\delta_w(\omega^k)\,.
\end{equation}

The optimization variables are the probabilities~$p^w_k$ (hence the
$\alpha_{i,j}$) and the
service rates~$\mu_i$, 
which appear in
the objective function in \Eq{obj-generic} via the CDF in 
\Eq{cdf-s-generic}. 
Ingress and egress rates~$\Lambda_i^k$ and~$M_i^k$ are input
parameters, as are the maximum service rates~$\mu^{\max}_i$ and the
target flow travel times~$\omega^k$. The per-queue arrival rates~$\lambda_i$
are auxiliary variables, 
whose dependency on the decision variables is specified in 
\Eq{lambda-p}.
Furthermore, we need to impose the constraints in \Eq{mu-max}  and \Eq{pwk-sum-one}.

{\bf Reconstructing the $\alpha$-values.}  
Given the optimal~$p_w^k$ values, the~$\alpha_{i,j}$ variables can be
recovered by solving a system of equations of the type of \Eq{pwk}, where
the $\alpha_{i,j}$-values are the unknown and the $p_w^k$-values are
given. The system can be linearized by taking the logarithm of all
variables, i.e., re-writing \Eq{pwk} as:  
$\log p_w^k=\sum_{n=2}^{|w|}\log\alpha_{w[n-1],w[n]}$, $\forall w\in\Wc$.

The fact that the system has a unique solution is ensured by the 
proposition below; the intuition behind it is that the number
of paths grows faster than the number of junctions, hence, there are
more equations than variables.  Thus, if it
  exists, the solution  is unique.

Solving the problem and finding the optimal value for the $p_w^k$~values could, {\em in principle}, be attempted with commercial, off-the-shelf solvers like CPLEX or
Gurobi, at least when a closed-form expression of \Eq{obj-generic} exists. However, numerical solvers encounter significant numerical
difficulties when dealing with the objective \Eq{obj-generic}.  Indeed, coefficients therein usually
contain a ratio of products of $\mu_i-\lambda_i$~terms: a small variation in any of the terms can change the sign of the whole coefficient and/or significantly alter its (absolute) value. The issue is compounded by the fact that such values, which can be very large, are then multiplied in \Eq{prob-k} by probabilities~$p_w^k$, which could be very small and/or be varied by very small quantities.

To avoid these issues, as well as to leverage problem-specific insights to obtain shorter solution times, we develop instead our own algorithm to solve the problem, as set forth next.

\section{The BH algorithm}
\label{sec:algo}

Our algorithm, named {\em bottleneck-hunting} (BH),  solves the problem specified in \Sec{problem} with the same {\em worst-case} complexity and convergence properties of gradient-based alternatives, but with faster {\em average-case} performance. Such a faster convergence is achieved by leveraging problem-specific knowledge and information in order to reduce the number of solutions to try out, hence, of algorithm iterations.

\begin{algorithm}[b]
\caption{The bottleneck-hunting (BH) algorithm\label{alg:bh}}
\begin{algorithmic}[1]

\State{$\phi\gets\phi_0$} \label{line:init-phi}
\While{$\textbf{true}$}
\State{$\texttt{CQ}\gets\{q_i\in\Qc\colon\exists w_1,w_2\in\Wc\colon
  q_i\in w_1\wedge q_i\in w_2 \wedge (\mu_i-\lambda_i) -\min_{q_j\neq q_i\in w_1} (\mu_j-\lambda_j) \leq\phi\Lambda^{\kappa(w_2)}\}$} \label{line:cq}

\State{$\texttt{CP}\gets\{w\in\Wc\colon w\cap\texttt{CQ}\not\equiv\emptyset\}$} \label{line:cp}
\State{$\texttt{FA}\gets\{k\in\Kc\colon\exists w\in\Wc\setminus\texttt{CP}\colon \kappa(w)=k\}$} \label{line:fa-ncp}
\If{$\texttt{FA}\equiv\emptyset$} \label{line:fa-check}
\State{$\texttt{FA}\gets\Kc$} \label{line:fa-all}
\EndIf

\State{$k^\star\gets\arg\max_{k\in\texttt{FA}} \delta^k(\omega^k)$} \label{line:kstar}
\State{$w^\star\gets\arg\max_{w\in\Wc\colon \kappa(w)=k^\star} \delta_w(\omega^{k^\star})$} \label{line:wstar}
\State{$w'\gets\arg\min_{w\in\Wc\colon\kappa(w)=k^\star} \max_{q_i\in w}(\mu_i-\lambda_i)$} \label{line:wprime}

\If{$\textsf{does\_improve}(k^\star,w^\star,w',\phi)$} \label{line:improves}
\State{$p_{w^\star}^{k^\star}\gets p_{w^\star}^{k^\star}-\phi$} \label{line:pminus}
\State{$p_{w'}^{k^\star}\gets p_{w'}^{k^\star}+\phi$} \label{line:pplus}
\Else
\State{$\phi\gets\frac{\phi}{2}$} \label{line:halfphi}
\If{$\phi<\phi^{\min}$} \label{line:phimax}
\State{$\textbf{return}$} \label{line:break}
\EndIf
\EndIf

\EndWhile
\end{algorithmic}
\end{algorithm}

With reference to \Alg{bh}, at every iteration, BH moves a
fraction~$\phi$ of flow~$k^\star$'s traffic from path~$w^\star$ to
path~$w'$. The fraction~$\phi$ changes across iterations, and is
initialized (\Line{init-phi}) to a value~$\phi_0$. Then, at every
iteration, BH identifies (\Line{cq}) a set \path{CQ} of {\em critical
  queues}, that is, queues that: (i) belong to two paths~$w_1$
and~$w_2$; (ii) are not the most loaded queue in~$w_1$; and, (iii)
 increasing their load by a fraction~$\phi$ of the
 traffic~$\Lambda^{\kappa(w_2)}$ of flow~$\kappa(w_2)$ renders it
 the most loaded queue in~$w_1$.
It follows that the algorithm will try to avoid routing additional traffic on critical queues in \path{CQ} if possible. Based on \path{CQ}, a set \path{CP} of {\em critical paths}, i.e., paths containing at least one critical queue, is identified in \Line{cp}.

Next, BH identifies the set \path{FA} of flows that can be acted
upon. Flows using at least one non-critical path are tried first
(\Line{fa-ncp}); if no such flow exists (\Line{fa-check}), then
\path{FA} is extended to include all flows in~$\Kc$
(\Line{fa-all}). The flow~$k^\star$ to act upon is chosen in
\Line{kstar}: considering the min-max nature of objective
\Eq{obj-generic}, $k^\star$~is the flow for which the summation in
\Eq{obj-generic} is largest. For the same reason, the path~$w^\star$
to {\em remove} vehicles from is chosen as the most loaded one, hence,
the one associated with the largest term in \Eq{obj-generic}. Similarly, the path~$w'$ to {\em add} traffic to is selected (\Line{wprime}) as the least-loaded one among those used by~$k^\star$; note that this implies choosing a non-critical path if such paths exist.

In \Line{improves}, BH calls the function {\sf does\_improve}, which
recomputes \Eq{obj-generic} and checks whether it improves by moving a
fraction~$\phi$ of $k^\star$'s vehicles from path~$w^\star$ to
path~$w'$. 
Indeed, the objective may not improve if the current value of~$\phi$
is too high, i.e.,  moving a $\phi$ fraction of
traffic increases the traffic intensity on $w'$ too much. In this case, such 
action should be performed at a later iteration, when $\phi$ will be
smaller (\Line{halfphi}). If the objective improves, 
the $p_w^k$~variables are updated accordingly, in
\Line{pminus}--\Line{pplus}. The algorithm terminates when $\phi$~drops below the minimum value~$\phi^{\min}$ (\Line{phimax}).

\begin{figure}
\centering
\includegraphics[width=1\columnwidth]{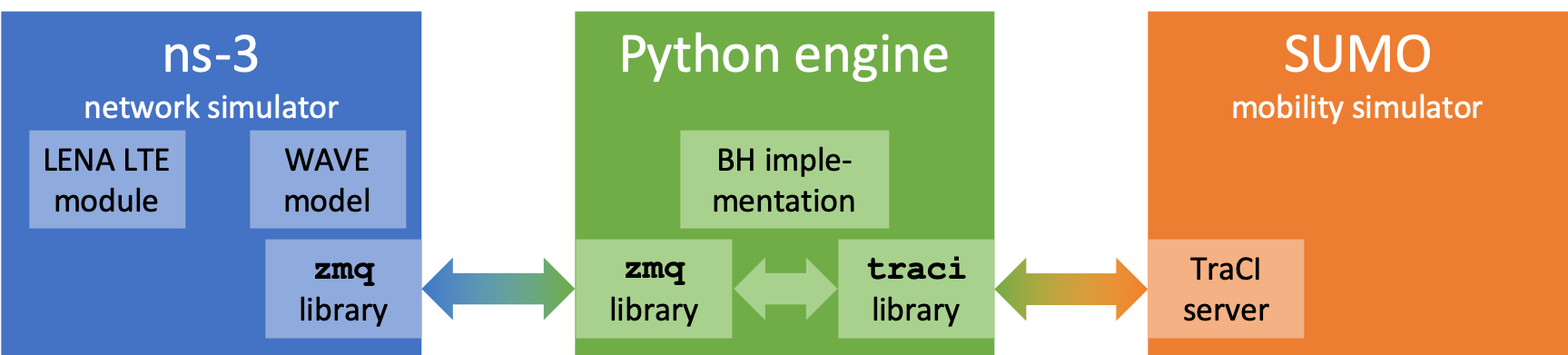}
\caption{Our validation framework, integrating the ns-3 network simulator,
a Python engine including the BH implementation, and the SUMO mobility simulator. 
\label{fig:validation}
\vspace{-4mm}
} 
\end{figure}

\begin{figure}
\centering
\includegraphics[width=.8\columnwidth]{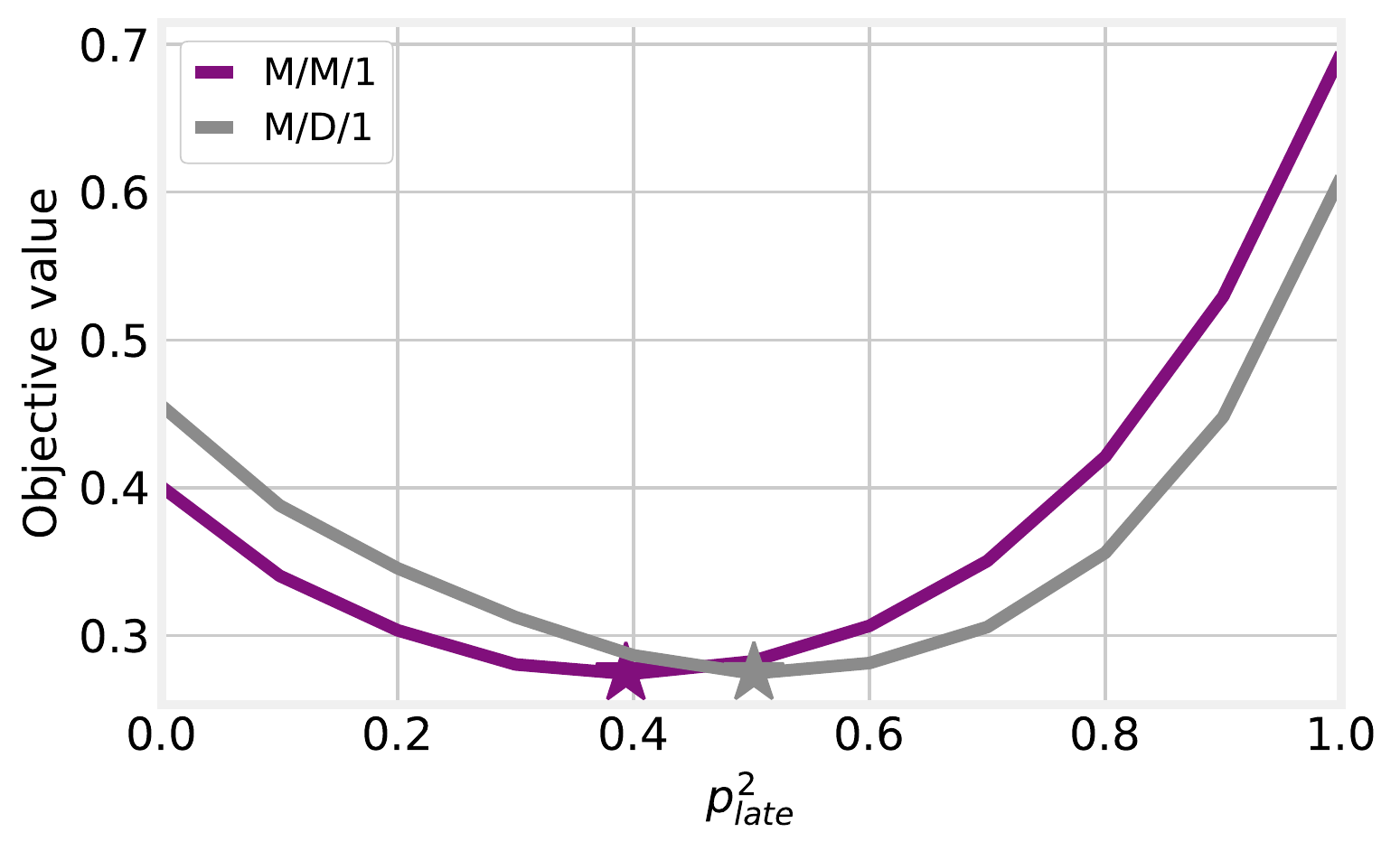}
\caption{
Lane change application: value of the objective \Eq{obj-generic} as a function of~$p^2_{\text{late}}$, when road segments are modeled as M/M/1 (purple) or M/D/1 (gray) queues.
    \label{fig:small-comparison}
    \vspace{-4mm}
} 
\end{figure}

\begin{figure*}
\centering
\includegraphics[width=.32\textwidth]{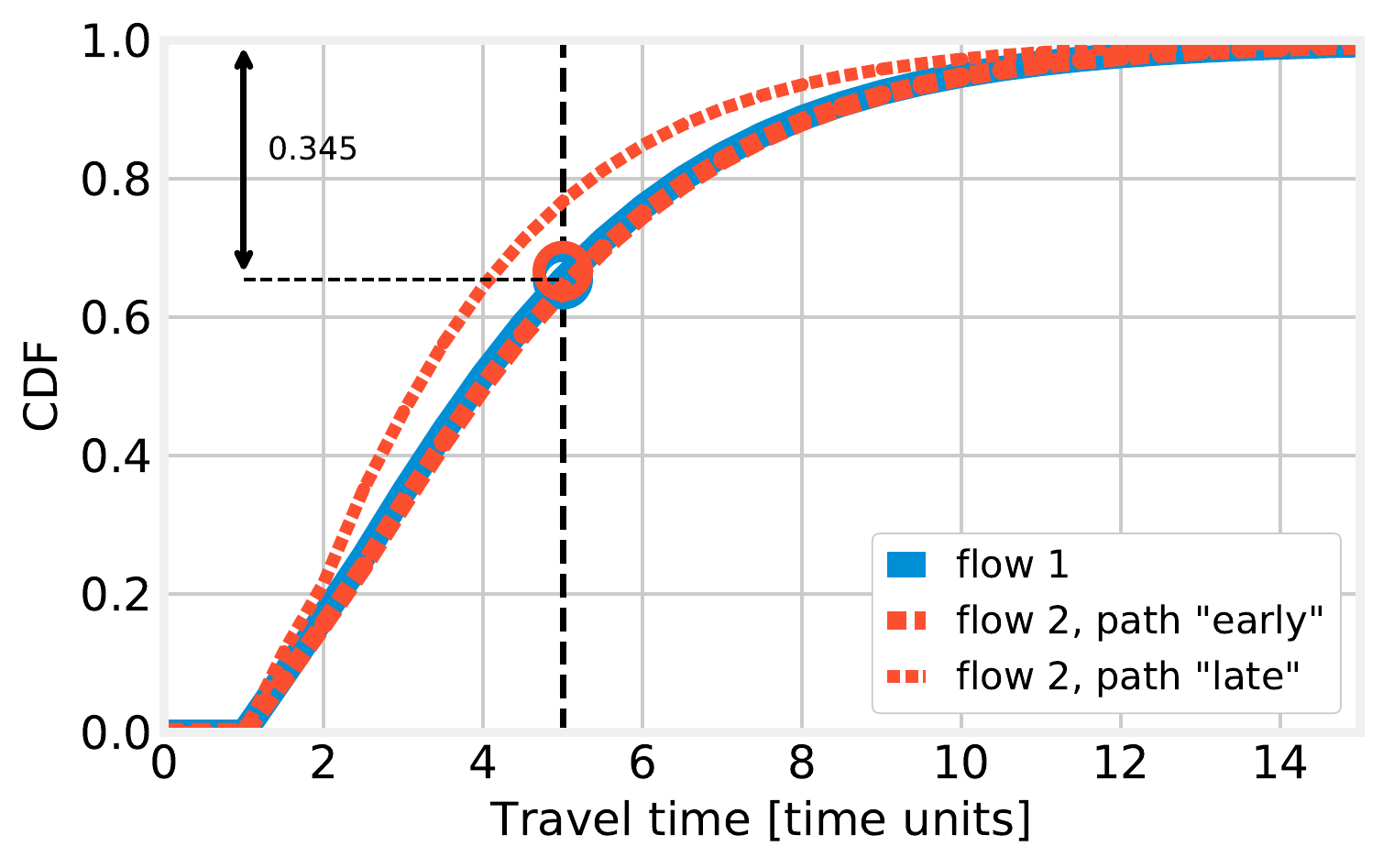}
\hspace{1mm}
\includegraphics[width=.32\textwidth]{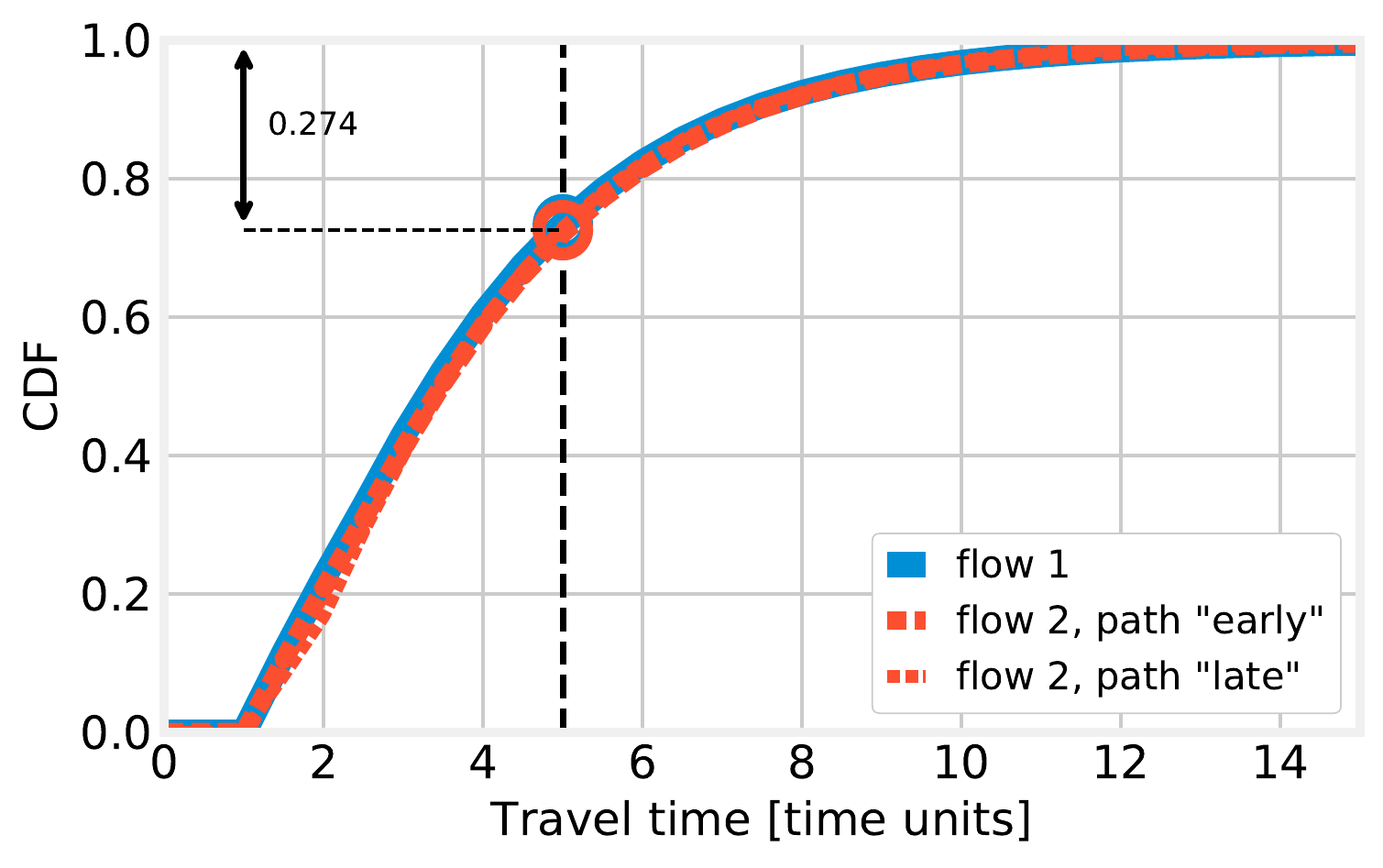}
\hspace{1mm}
\includegraphics[width=.32\textwidth]{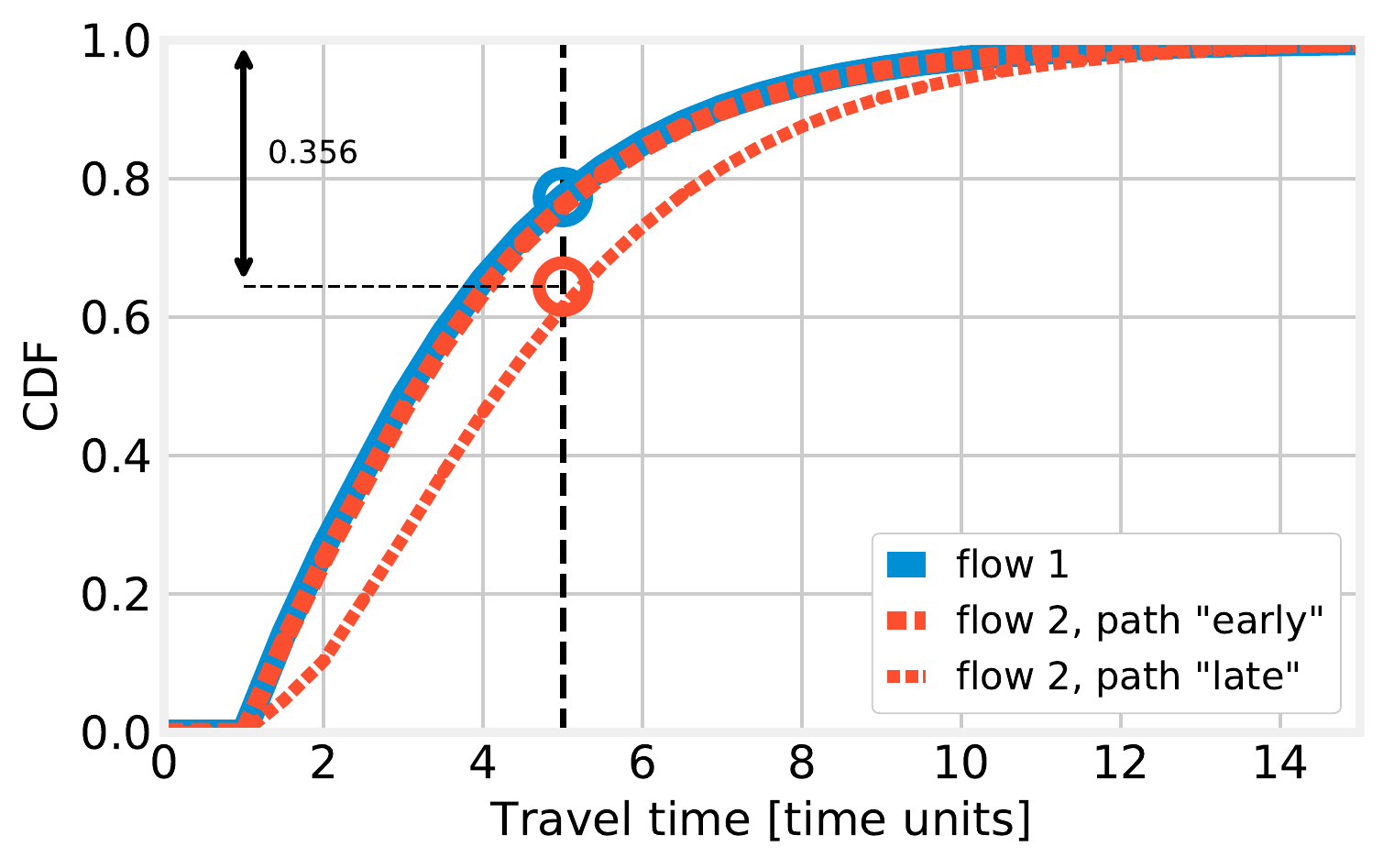}
\caption{
Small-scale scenario, lane-change,  M/D/1 queues: CDF of the per-path travel time when~$p^2_{\text{late}}=0.1$ (left), $p^2_{\text{late}}$~takes its optimal value (center), and $p^2_{\text{late}}=0.9$ (right).
    \label{fig:small-cdfs-md1}
} 
\end{figure*}

\section{Validation Methodology}
\label{sec:validation}

 To assess its performance, we integrate BH within a complete validation
environment, as represented in
\Fig{validation}. Policies summarized by
the~$p_w^k$ and $\mu_i$ variables are defined by BH, implemented within a 
Python engine. Such decisions are then relayed to the ns-3 network simulator,
which is in charge of:
   (i) simulating the network traffic generated by vehicles and by the infrastructure, and (ii) based on
   the most recent CAMs,  implementing
   the edge-defined policy while minimizing the disruption for the
   vehicles. 

In the case of the lane-change service, upon receiving a DENM, vehicles engage their neighbors in a
communication following the  protocol in \cite{autonet30-commag}, to
coordinate their  manoeuvers and avoid collisions. 
As a consequence, the target vehicle as well as its 
neighbors may vary their trajectory, and such a variation may have cascade effects on the mobility
of other vehicles. All  mobility, in both  the lane-change and the 
navigation case, is simulated via
SUMO. 
Based on the SUMO simulation, the position of each vehicle
is then updated within ns-3.

The mobility information is relayed between ns-3 and SUMO through the Python
engine and the TraCI Python library. 
The ns3 simulator and the Python
engine interact  through the \path{zmq} message-passing framework, using the client libraries available for both Python and C++. The communication between SUMO  and
the Python engine, instead, takes place through the TraCI protocol.

In SUMO, flows include a mixture of different vehicle types, namely cars (SUMO class \path{passenger}, 85\% of all vehicles), trucks (class \path{truck}, 10\% of vehicles), and buses (class \path{coach}, 5\% of vehicles). Their target speed, acceleration, and driving aggressiveness values are left to the SUMO default, subject to a global speed limit of 50~km/h, as it is common in urban areas. The simulation step size of SUMO, which also determines the frequency of position updates in ns-3, is set to 10~ms.

In ns-3, LTE (provided by the LENA module) is used for the
communication between vehicles and infrastructure, while WAVE
(in the default ns-3 distribution) is used for
V2V communication, including that required for
lane change. CAMs and DENMs are encoded as foreseen by the ETSI 302.637
standard for ITS. 
Upon receiving a DENM, vehicles take action immediately, which
corresponds to the case of autonomous vehicles. Human reaction times,
usually quantified in 1\,s~\cite{noi-vtm}, could be easily accounted for.

\section{Numerical Results}
\label{sec:results}

We now demonstrate how our system model and the BH algorithm can be
exploited, by considering a lane-change service in the small-scale
scenario described in \Fig{archi} and modelled in \Fig{scenario}. Therein, there are two incoming
flows, namely, flows~1 and~2: the former is associated with one path
only, the latter with two paths. These two paths are called \path{early}
and \path{late}, referring to the fact that vehicles turn north
(respectively) after the first road segment (i.e., $q_2$), or after the
second one (i.e., $q_4$). We set the normalized\footnote{Rates are
  normalized to the arrival rate of flow 1.} incoming rate to
$\Lambda^k=1$~for both flows, and the
maximum normalized service rate to $\mu^{\max}_i=3$ for all road segments, except for $q_4$ that, owing to
the fact that vehicles therein need to slow down, has a 
  maximum  normalized service rate $\mu_4^{\max}=1.5$. Also, we set
  $\omega^k$ to 5 time units for both flows. 
In such a  scenario, the optimal values of $\mu_i$ coincide with
$\mu^{\max}_i$ for  all road segments, thus 
the decisions to make are summarized by the
variable~$p^2_{\text{late}}$ (indeed, flow~1 only has one path and
$p^2_{\text{early}}=1-p^2_{\text{late}}$). 

The first, high-level question we seek to answer concerns the
relationship between the variable~$p^2_{\text{late}}$ and the value of
the objective function \Eq{obj-generic}. \Fig{small-comparison} shows
that it is advantageous to split the vehicles of flow~2 more or less
evenly between its two possible paths.

The third prominent message conveyed by \Fig{small-comparison} is about
the flexibility of our approach:  the purple
curve is obtained by modeling road segments as M/M/1 queues, thus
using closed-form expressions, while
the gray curve is obtained by using M/D/1 queues. Since there is
{\em no closed-form expression} of the sojourn time distribution in
M/D/1 queues, we 
implemented the approximate formula presented in~\cite{md1-times}, and
solved numerically integrals and convolutions. In spite of
the very significant differences with respect to the M/M/1 case, our
approach and BH  {\em work with no changes} in both
cases: our solution strategy does not depend on any specific service time distribution, and does not require such a distribution to have a closed-form expression.

Still for  the M/D/1 case, \Fig{small-cdfs-md1} shows the distribution of the per-path travel times, along with a graphical interpretation of the~$\omega^k$, $\delta_w(\omega^k)$, and~$\delta^k(\omega^k)$ quantities.
Each curve corresponds to a path, and its color represents the flow
each path belongs to (namely, blue: flow 1, red: flow 2). Firstly, we
observe  that, as the intuition would suggest, a higher value
of~$p^2_{\text{late}}$, i.e., sending more vehicles to path
\path{late} of flow~2, results in longer travel times for that path,
and shorter ones for path \path{early}. Secondly, the black vertical line in all
plots corresponds to  the value of the target travel time $\omega^k$;
ideally, one would like all CDFs to be at the left of such a line. 
 The intersections
between the CDFs and the vertical line represent the
probability that  vehicles belonging to a flow take a {\em path} whose travel time
does not exceed~$\omega^k$, i.e., $(1-\delta_w(\omega^k))$. 
Instead, the circles noted on the plots represent the {\em flow}-wise performance~$\delta^k(\omega^k)$:
as specified in \Eq{prob-k}, such a
quantity is defined as a weighted sum of the $\delta_w(\omega^k)$
values, thus, $\delta^2(\omega^2)$ is close
to~$\delta_\text{early}(\omega^2)$ when most vehicles take path
\path{early} (left plot), and close
to~$\delta_\text{late}(\omega^2)$ in the opposite case (right plot). 

The lowest circle, and the numerical values reported in the plots, correspond to the largest value
of~$\delta^k(\omega^k)$, i.e., the value of the objective function in \Eq{obj-generic}: intuitively, optimizing \Eq{obj-generic}
corresponds to pushing up  such a circle as high as possible. 
Note that the flow with the highest~$\delta^k(\omega^k)$ changes for different values of~$p^2_\text{late}$: it is flow~1 in the left plot, and  flow~2 in the right one.
In the center plot, corresponding to the optimal solution, all paths
exhibit roughly the same travel time distribution. This is consistent with the intuition we leverage for the design of the BH algorithm: similar path travel times result in better performance.

\section{Related work}
\label{sec:relwork}

Queues have been consistently and effectively used to represent
vehicular traffic~\cite{queue-survey}. As early as 1995,
\cite{alfa1995modelling}~argued that vehicles arrive at intersection,
either individually or in batches, following a Poisson process.
The later work~\cite{van2006empirical} leveraged empirical validation to argue for a more general, M/G/1 model in non-congested scenarios. In both cases, individual lanes are each associated with a queue. Vehicular mobility is modeled through M/M/1 queues also in several recent works about edge and cloud computing in vehicular networks, including~\cite{zhou2018begin,ning2019vehicular,zhao2018joint}.

Other works, e.g.,~\cite{mm1-indiani},
focus on specific aspects of vehicular traffic, such as modeling the
delay of indisciplined traffic:
M/M/1 models are found to accurately represent real-world
conditions. 
Consistently, the study in~\cite{carlo-secon12}, based on
realistic simulations, finds M/M/1 to well represent travel time on
individual road segments.
Note that part of the popularity of Markovian service times is due to their
ability to capture the effect of heterogeneous traffic (e.g., cars, trucks,...)~\cite{mm1-indiani}, proceeding at different speeds.

Batch arrival and departure processes are the modeling tool of choice
when traffic signals are involved,
e.g.,~\cite{timmerman2019platoon,harahap2019modeling}. In
all cases, arrivals are assumed to be Markovian in nature, while
service times are either Markovian or deterministic.

Situations where vehicles from multiple flows have to {\em merge} into
one lane has been identified as a major cause of
congestion. Accordingly, several research efforts focus on
modeling~\cite{kanagaraj2010modeling} the merging behavior of
vehicles.
Recent efforts such as~\cite{zhou2016impact} deal with merging strategies for autonomous vehicles, and the benefits coming from cooperation between them. Finally, \cite{bushnell1970merging}~envisions a centralized controller for merge manoeuvers, accounting for the impact of local merging behaviors on the global traffic conditions.


\section{Conclusions}
\label{sec:conclusion}

We considered assisted driving services for connected vehicles, aiming
at optimizing the traffic flows. We first introduced a system
architecture that integrates the distributed vehicular network with the 
edge network, thus enabling  
edge-controlled traffic flows.  Then we developed a queue-based model
describing arbitrary road topologies and the behavior of the vehicles
traveling through them. Through such a model, we derived the
distribution of  the vehicle travel times and used it to formulate
the problem of optimizing such times.
We then devised 
 the BH algorithm, which provides optimal policies in linear time
 without requiring a closed-form expression for the distribution of
 the queue service time.
Our results, derived through a comprehensive and realistic simulation
framework, confirm that the policies formulated by BH result in much
shorter travel times than distributed decisions.

\section*{Acknowledgments}
This work was supported by TIM through the contract ``Performance Analysis of Edge Solutions for Automotive Applications''.

\bibliographystyle{IEEEtran}
\bibliography{refs_short}

\begin{thebibliography}{10}
\providecommand{\url}[1]{#1}
\csname url@samestyle\endcsname
\providecommand{\newblock}{\relax}
\providecommand{\bibinfo}[2]{#2}
\providecommand{\BIBentrySTDinterwordspacing}{\spaceskip=0pt\relax}
\providecommand{\BIBentryALTinterwordstretchfactor}{4}
\providecommand{\BIBentryALTinterwordspacing}{\spaceskip=\fontdimen2\font plus
\BIBentryALTinterwordstretchfactor\fontdimen3\font minus
  \fontdimen4\font\relax}
\providecommand{\BIBforeignlanguage}[2]{{%
\expandafter\ifx\csname l@#1\endcsname\relax
\typeout{** WARNING: IEEEtran.bst: No hyphenation pattern has been}%
\typeout{** loaded for the language `#1'. Using the pattern for}%
\typeout{** the default language instead.}%
\else
\language=\csname l@#1\endcsname
\fi
#2}}
\providecommand{\BIBdecl}{\relax}
\BIBdecl

\bibitem{who-roads}
{World Health Organization}. {Global status report on road safety}.

\bibitem{zhou2016impact}
M.~Zhou \emph{et~al.}, ``On the impact of cooperative autonomous vehicles in
  improving freeway merging: a modified intelligent driver model-based
  approach,'' \emph{IEEE Trans. on ITS}, 2016.

\bibitem{lehoczky1972traffic}
J.~Lehoczky, ``Traffic intersection control and zero-switch queues under
  conditions of markov chain dependence input,'' \emph{Journal of Applied
  Probability}, 1972.

\bibitem{dunne1967traffic}
M.~C. Dunne, ``Traffic delay at a signalized intersection with binomial
  arrivals,'' \emph{INFORMS Transportation Science}, 1967.

\bibitem{autonet30-commag}
L.~Hobert \emph{et~al.}, ``Enhancements of v2x communication in support of
  cooperative autonomous driving,'' \emph{IEEE Comm. Mag.}, 2015.

\bibitem{5gcar}
M.~Fallgren \emph{et~al.}, ``Fifth-generation technologies for the connected
  car: Capable systems for vehicle-to-anything communications,'' \emph{IEEE
  Veh. Tech. Mag.}, 2018.

\bibitem{5gt}
A.~de~la Oliva \emph{et~al.}, ``5g-transformer: Slicing and orchestrating
  transport networks for industry verticals,'' \emph{IEEE Comm. Mag.}, 2018.

\bibitem{etsi-2019}
{ETSI}, ``{ETSI} 103 299 - v2.1.1,'' Tech. Rep., 2019.

\bibitem{noi-vtm}
M.~Malinverno \emph{et~al.}, ``Mec-based collision avoidance for vehicles and
  vulnerable users,'' \emph{ArXiv preprint 1911.05299}, 2019.

\bibitem{md1-times}
V.~Iversen \emph{et~al.}, ``{Waiting time distribution in M/D/1 queueing
  systems},'' \emph{IEEE Electronics Letters}, 2000.

\bibitem{queue-survey}
T.~Van~Woensel \emph{et~al.}, ``Modeling traffic flows with queueing models: A
  review,'' \emph{Asia-Pacific Journal of Operational Research (APJOR)}, 2007.

\bibitem{alfa1995modelling}
A.~S. Alfa \emph{et~al.}, ``Modelling vehicular traffic using the discrete time
  markovian arrival process,'' \emph{INFORMS Transportation Science}, 1995.

\bibitem{van2006empirical}
T.~Van~Woensel \emph{et~al.}, ``Empirical validation of a queueing approach to
  uninterrupted traffic flows,'' \emph{Springer 4OR}, 2006.

\bibitem{zhou2018begin}
Z.~Zhou \emph{et~al.}, ``Begin: Big data enabled energy-efficient vehicular
  edge computing,'' \emph{IEEE Comm. Mag.}, 2018.

\bibitem{ning2019vehicular}
Z.~Ning \emph{et~al.}, ``Vehicular fog computing: Enabling real-time traffic
  management for smart cities,'' \emph{IEEE Wireless Comm.}, 2019.

\bibitem{zhao2018joint}
L.~Zhao \emph{et~al.}, ``Joint optimization of communication and traffic
  efficiency in vehicular networks,'' \emph{IEEE Trans. on Veh. Tech.}, 2018.

\bibitem{mm1-indiani}
S.~Mukhopadhyay \emph{et~al.}, ``Approximate mean delay analysis for a
  signalized intersection with indisciplined traffic,'' \emph{IEEE Trans. on
  ITS}, 2017.

\bibitem{carlo-secon12}
C.~{Borgiattino} \emph{et~al.}, ``Modelling realistic vehicle traffic flows,''
  in \emph{IEEE SECON VCSC Workshop}, 2012.

\bibitem{timmerman2019platoon}
R.~Timmerman \emph{et~al.}, ``Platoon forming algorithms for intelligent street
  intersections,'' in \emph{Mathematics Applied in Transport and Traffic
  Systems}, 2019.

\bibitem{harahap2019modeling}
E.~Harahap \emph{et~al.}, ``Modeling and simulation of queue waiting time at
  traffic light intersection,'' in \emph{IOP Journal of Physics: Conference
  Series}, 2019.

\bibitem{kanagaraj2010modeling}
V.~Kanagaraj \emph{et~al.}, ``Modeling vehicular merging behavior under
  heterogeneous traffic conditions,'' \emph{SAGE Transportation Research
  Record}, 2010.

\bibitem{bushnell1970merging}
D.~Bushnell, ``A merging control system for the urban freeway,'' \emph{IEEE
  Trans. on Veh. Tech.}, 1970.

\end{thebibliography}

\end{document}